\RequirePackage{lineno}
\documentclass[showpacs,preprintnumbers,amsmath,amssymb]{revtex4}
\usepackage{graphicx}
\usepackage{dcolumn}
\usepackage{lineno}
\usepackage{bm}
\usepackage{color}

\newcommand{ \be }{\begin{linenomath*}\begin{equation}}
\newcommand{ \ee }{\end{equation}\end{linenomath*}}
\newcommand{ \bea }{\begin{linenomath*}\begin{eqnarray}}
\newcommand{ \eea }{\end{eqnarray}\end{linenomath*}}
\newcommand{ \la }{\langle}

\newcommand{ \ra }{\rangle}

\usepackage{color}

\begin{document}
\begin{setpagewiselinenumbers}
\begin{linenumbers}
\title{Longitudinal dependence of two-particle momentum correlations from NEXSPHERIO model}
\author{Monika Sharma, Claude Pruneau, and Sean Gavin}
\affiliation{Department of Physics and Astronomy, Wayne State University, Detroit, MI 48201 USA}
\author{Jun Takahashi,  R. Derradi de Souza}
\affiliation{Universidade Estaudal de Campinas, S‹o Paulo, 13083-970, Brazil}
\author{T. Kodama}
\affiliation{Universidade Federal do Rio de Janeiro, Rio de Janeiro, 21945-970, Brazil}
Version 2.4, Aug 31, 2011
\begin{linenomath*}
\begin{abstract}
The rapidity dependence of two-particle momentum correlations can be used to probe the viscosity of the liquid produced in heavy nuclei collisions at RHIC. We reexamine this probe in light of the recent experimental analyses of the azimuthal-angle dependence of number correlations, which demonstrate the importance of  initial state fluctuations propagated by hydrodynamic flow in these correlations. The NEXSPHERIO model combines fluctuating initial conditions with viscosity-free hydrodynamic evolution and, indeed, has been shown to describe azimuthal correlations.  We use this model to compute  the number density correlation $R_{2}$ and the momentum current correlation function {\it C}, at low transverse momentum in  Au+Au collisions at $\sqrt{s_{NN}} = $~200 GeV. The correlation function {\it C} is sensitive to details of the collision dynamics. Its longitudinal width is expected to broaden under the influence of viscous effects and narrow in the presence of sizable radial flow. While the NEXSPHERIO model qualitatively describes the emergence of a near-side ridge-like structure for both the $R_2$ and {\it C} observables, we find that it predicts a longitudinal narrowing of the near side peak of these correlation functions for increasing number of participants in  contrast with recent observations by the STAR Collaboration of a significant broadening in most central collisions relative to peripheral collisions.
\end{abstract}
\end{linenomath*}


\maketitle
\setcounter{page}{1}

\section{Introduction}
\label{sec:introduction}

Measurements of elliptic flow indicate that strong collective flow is achieved in nuclear collisions at RHIC \cite{Ref1,Ref2,Ref3,Ref4,Ref5}; see also the reviews \cite{Voloshin:2008dg,Sorensen:2009cz}. Hydrodynamic models describe many features of the experimental data, indicating that the produced medium behaves as a nearly ideal fluid, i.e. a fluid endowed with a very small shear viscosity per unit of entropy, $\eta/s$. Considerable effort is devoted to using elliptic flow data to measure  $\eta/s$; see, e.g., \cite{Shen:2011eg,Lacey:2010fe}, and Refs.\ therein. However, the interpretation of such measurements has many sources of theoretical uncertainty. The primary uncertainties are: initial conditions, event-by-event fluctuations, and freeze-out \cite{Kodama}. These effects are somewhat outside the scope of hydrodynamics and further theoretical work is required to elaborate a systematic scheme  to compute them with quantifiable uncertainties. It is therefore important to explore independent methods to determine $\eta/s$.

One such alternative method was proposed by Gavin {\it et al.} based on two-particle transverse momentum correlations \cite{Gavin}.  This novel technique relies on the fact that viscous forces impart momentum transfers between different fluid cells. These momentum transfers in turn modify the relative rapidity of correlated particles produced by these  cells thereby leading to a longitudinal broadening of transverse momentum two-particle correlation functions. The longitudinal profile of this correlation function, hereafter called {\it C}, is expected to broaden  with increasing collision system life-time, and for increasing number of collision participants. The measurement strategy proposed by authors of   \cite{Gavin} is to measure the $p_T$ correlation function ${\it C}$ as a function of collision centrality and seek evidence for its longitudinal broadening in most central collisions. Given its explicit dependence on particle momenta, the correlation function ${\it C}$ is sensitive to momentum current correlations and as such constitutes a more suitable observable to measure viscosity than simple two particle (number) correlation functions.

The STAR experiment recently carried an analysis of a differential version of the $p_T$ correlation function {\it C}.  STAR reported that the near-side of the correlation function broadens progressively with increasing collision centrality \cite{msharma1}. The broadening is considerable but translates nonetheless in a rather modest $\eta/s$ consistent with values deduced from comparison of hydrodynamic calculations with $v_2$ measurements. While, this agreement is encouraging,  one must however exercise caution in the interpretation of the broadening of the measured correlation function. Indeed, one must assess whether other mechanisms could produce or perhaps mask broadening. Reaction mechanisms and effects of particular interest include two-particle kinematical focusing associated with radial flow, jet production, and more specifically jet quenching, flux tubes, fluctuations, anisotropic flow, resonances, and momentum conservation. One must also acknowledge that the structure of two-particle correlations observed is not fully elucidated. While differential correlation functions have been previously measured by the STAR \cite{Ref7}, PHENIX \cite{Ref8} and PHOBOS \cite{Ref9} experiments, a significant interest lies in studying diverse correlation functions which provide better understanding and sensitivity to various reaction mechanisms. Dihadron correlations have previously been studied containing a high-$p_{T}$ ``trigger" particle and a low-$p_{T}$ ``associated" particle. Measurements of near-side enhanced yield - long range correlations - also known as the ``Ridge" were reported by the STAR collaboration \cite{Ref10,Ref11}. Some theoretical models attribute the ridge to jet-medium interactions: particles from jet fragmentations in QCD vacuum result in a peak at $\Delta \eta \sim 0$ and those affected by the medium are diffused broadly in pseudorapidity, forming a ridge-like structure \cite{Ref12,Ref12:a,Ref12:b,Ref12:c,Ref12:d,Ref12:e,Ref12:f,Ref12:g}. Other plausible explanations of the ridge formation are given by transverse radial flow models where kinematic focussing of clusters, strings, or color flux tubes is induced by increasing radial flow with centrality \cite{Ref13}. Transverse radial expansion produces strong position-momentum correlations that leads to characteristic rapidity, $p_{T}$ and azimuthal correlations among the produced particles. Another interesting and promising interpretation of the near-side ridge observed in both triggered and untriggered  correlation functions is based on the notion of initial density fluctuations. Initial density fluctuations in the coordinate space shall lead to fluctuations in the number of locally produced particles. These fluctuations translate, under the influence of hydrodynamical radial expansion, into long range pseudorapidity two-particle correlations. Such long range correlations are well demonstrated in the NEXSPHERIO simulations reported by Takahashi {\it et al.}\cite{Ref17}.  
Other studies based on hydrodynamical and hybrid models have essentially confirmed this important observation  \cite{Petersen:2009vx,Werner:2010aa,Holopainen:2010gz,Hirano:2009bd,Petersen:2008dd}. Also note that much work is being done to quantatively understand the connection between initial collision system spatial anisotropies and final state flow coefficients, $v_n$, of orders n = 1 through 6 \cite{Sorensen:2011hm,Alver:2010gr,Petersen:2011fp,Petersen:2010cw,Armesto:2006bv}, and to determine the fluid viscosity (see for instance 
\cite{Song:2011qa,Romatschke:2007mq}).  NEXSPHERIO describes many features of the measured correlations, particularly their azimuthal dependence. It is thus of interest to extend the work of these authors and consider predictions of NEXSPHERIO for transverse momentum two-particle correlations.  

The NEXSPHERIO model, the correlation observable of interest, and the analysis technique used in this work are described in the next section. Results are presented in sec. \ref{sec:results}, and conclusions summarized in sec. \ref{sec:conclusions}.

\section{NEXSPHERIO Model and Analysis Technique}
\label{sec:nexspherio}

NEXSPHERIO \cite{Ref18,Ref19} combines the event generator NEXUS \cite{Ref15} and the hydrodynamical code SPHERIO \cite{Ref16} to simulate particle production in $A+A$ collisions. NEXUS provides for initial parton-parton interactions and energy deposition with lumpy initial conditions (ICs) specified in terms of number density, energy density and baryon density. SPHERIO is then used to simulate the hydrodynamical evolution in each NEXUS event, i.e thereby including the effect of IC fluctuations in the production of collective flow patterns and correlated particle production. Indeed, the recent studies with NEXSPHERIO show that hydrodynamic expansion of fluctuating ICs  lead to the formation of near-side ridge-like structure in $\Delta\eta$ vs. $\Delta\phi$ correlation functions \cite{Ref17} while, by contrast, smooth initial conditions do not feature any near-side ridge-like structures. Furthermore, an interesting interpretation of the origin of the away-side double peaked structure of the two particle correlation function was also discussed in the context of the NEXSPHERIO lumpy initial condition in reference \cite{LUMPY}.

The transverse momentum two-particle correlation {\it C}  depends at a basic level on particle number fluctuations and should therefore exhibit the formation of a ridge, in NEXSPHERIO, for central collisions with lumpy initial conditions. {\it C} also depends on actual momentum-momentum correlations. Its magnitude and dependence on $\Delta\eta$ vs. $\Delta\phi$ should thus also be influenced by radial hydrodynamical flow brought about by lumpy initial density conditions. However, given SPHERIO is built as an ideal hydrodynamical transport code, there should be no viscous effects: no broadening of the correlation function should arise with increasing collision centrality. Any change in the correlation function profile and width should be due solely to viscous free expansion dynamics. 

Our analysis is based on a total of 500,000 simulated Au + Au collisions events  at $\sqrt{s_{NN}} = $~200 \mbox{GeV} produced with NEXSPHERIO ~\cite{Ref17}. Events were grouped in four collision centrality classes corresponding to 0-10\% (most central), 20-30\%, 40-60\% and 60-80\% (most peripheral) fractions of the total reaction cross-section. Modeling nuclear collisions in ideal hydrodynamics is limited to low $p_{T}$ ($\leq$ 2.0 \mbox{GeV}/{\it c}) because it is generally assumed that the equilibrium description fails at high $p_{T}$, where particle production may be dominated by hard processes. We therefore focus our analysis on bulk particles, i.e. particles produced in the range $p_T < 2.0$ \mbox{GeV}/{\it c}. This momentum selection is used for both particles in constructing 
the two particle correlation function $R_2$, and the momentum dependent correlation function, {\it C}, defined below, thereby yielding ``untriggered" distributions. We simulate the STAR experiment acceptance and further limit our simulations to particles produced in the pseudorapidity range $|\eta|<1.0$ and with $0.2< p_T < 2.0$ \mbox{GeV}/{\it c}.  

We carried an analysis of observables $R_{2}$ and {\it C} studied as a function of pseudorapidity ($\Delta\eta$) and azimuth($\Delta\phi$) differences of particle pairs. These correlation functions are defined as follows:
\begin{linenomath*}
\begin{eqnarray}
R_2(\Delta\eta,\Delta\phi) &=& \frac{ \int{\rho_{2}(p_{T,1},\eta_1,\phi_1,p_{T,2},\eta_2,\phi_2) \delta(\Delta\eta-\eta_1+\eta_2) \delta(\Delta\phi-\phi_1+\phi_2) dp_{T,1}d\eta_1 d\phi_1 dp_{T,2}d\eta_2 d\phi_2}}{\int{\rho_{1}(\eta_1,\phi_1)\rho_{1}(\eta_2,\phi_2) \delta(\Delta\eta-\eta_1+\eta_2) \delta(\Delta\phi-\phi_1+\phi_2) dp_{T,1} d\eta_1 d\phi_1 dp_{T,2} d\eta_2 d\phi_2}}-1\\ \nonumber
C(\Delta\eta,\Delta\phi)     &=&  \frac{ \int{\rho_{2}(p_{T,1},\eta_1,\phi_1,p_{T,2},\eta_2,\phi_2)p_{T,1}p_{T,2} \delta(\Delta\eta-\eta_1+\eta_2) \delta(\Delta\phi-\phi_1+\phi_2) dp_{T,1}d\eta_1 d\phi_1 dp_{T,2}d\eta_2 d\phi_2}}{\int{\rho_{1}(\eta_1,\phi_1)\rho_{1}(\eta_2,\phi_2) \delta(\Delta\eta-\eta_1+\eta_2) \delta(\Delta\phi-\phi_1+\phi_2) dp_{T,1}d\eta_1\phi_1 dp_{T,2}d\eta_2d\phi_2}} - \la p_{T}\ra_1  \la p_{T}\ra_2
\label{Eq1}
\end{eqnarray}
\end{linenomath*}
where $\rho_{1}=dN_1/dp_{T} d\eta d\phi$ and $\rho_{2}=dN_2/dp_{t,1} d\eta_1 d\phi_1dp_{t,2} d\eta_2 d\phi_2$ correspond to single and two-particle densities, respectively; $\left\langle {p_t } \right\rangle_k  \equiv \left\langle {\sum {p_{t,i } } } \right\rangle_k /\left\langle n \right\rangle_k$ is the average particle momentum. The  label {\it k=1,2} stands for particles being detected
at $(p_{t,k},\eta_k,\phi_k)$, and the brackets represent event ensemble averages.  The integrals are taken over $2\pi$ azimuth and in the same pseudorapidity range for both particles. The delta functions are used to restrict correlation yield contributions to relative azimuthal angle and pseudorapidity differences only.
The observable $R_2$ measures the correlation between the number of particles emitted at relative pseudorapidity $\Delta\eta$ and azimuthal angle difference $\Delta\phi$. It is a robust observable and is expected to scale inversely to the number of correlated particle sources in the absence of collective effects and rescattering of secondaries. The observable {\it C} amounts to a differential version of the integral correlation used by authors of~\cite{Gavin}. It is sensitive not only to the degree of correlation between produced particles but to the hardness of particle spectra, and correlations between the momentum of particles.

\section{Results}
\label{sec:results}

We present, in Fig.~\ref{CF:R2}, the correlation function  $R_{2}$  for (a) 60-80\% (most peripheral), (b) 40-60\%, (c) 20-30\% and (d) 0-10\% (most central) collision centralities. Our calculation of $R_2$ reproduces the features already reported in \cite{Ref17} for a `per trigger' correlation function. $R_2$ exhibits a somewhat narrow near-side peak (i.e. near $\Delta\eta\approx\Delta\phi\approx 0$) and an extended ridge-like  structure centered at $\Delta\phi\approx\pi$ in all collision centralities. The correlation function exhibits a strong $cos(2\Delta\phi)$ modulation associated with collective flow in mid-central collisions. This component is minimal in most central collisions but $R_2$ exhibits the build up of a narrow ridge at $\Delta\phi\sim 0$ which extends over the full $\Delta\eta$ range of the correlation function. 

Figure~\ref{CF:C} presents the correlation function  {\it C}  for the same four collision centrality bins  
used for $R_2$ in Fig. \ref{CF:R2}. {\it C} exhibits correlation shape dependencies on $\Delta\eta$ and $\Delta\phi$ qualitatively similar to those observed for $R_2$. It also has a narrow near-side peak at $\Delta\eta \approx \Delta\phi \approx 0$ and an extended $\Delta\eta$ ridge at $\Delta\phi\approx\pi$ in all collision centralities. Mid-central bins feature strong elliptic flow modulations and the most central bin exhibit a near-side peak atop an extended ridge. Also note that both $R_2$ and {\it C} exhibit an overall decrease of the correlation function strength  consistent with the increase in the average number of participant nucleons  and the decrease of the elliptic flow modulation from 60-80\% to 0-5\% collision centrality ranges. 

\begin{figure*}[h!]
\resizebox{\linewidth}{5.5cm}{\includegraphics[angle=90]{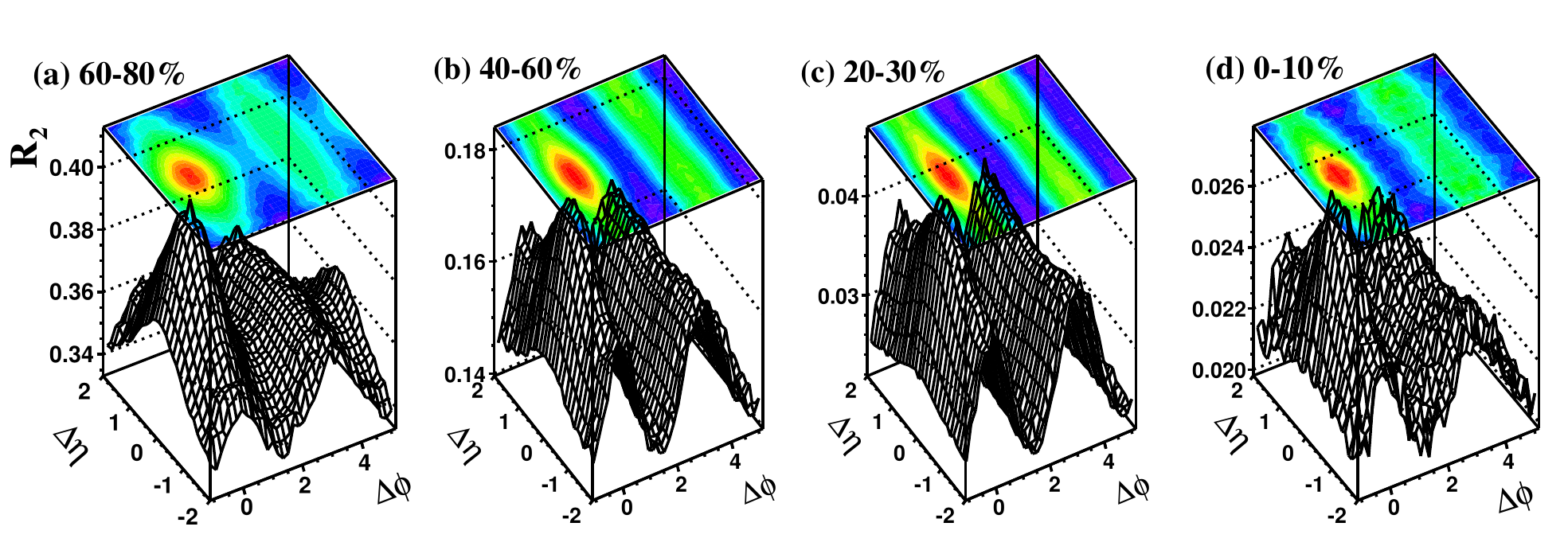}}
\caption[]{(Color online) Two-particle correlation function, $R_{2}$, plotted as a function of the particle pair pseudorapidity difference, $\Delta\eta$, and relative azimuthal angle, $\Delta\phi$ [rad], for (a) 60-80\%, (b) 40-60\%, (c) 20-30\% and (d) 0-10\% collision centralities in Au + Au interactions at $\sqrt{s_{NN}}$ = 200 \mbox{GeV} calculated with the NEXSPHERIO event generator.}
\label{CF:R2}
\end{figure*}

\begin{figure*}[h!]
\resizebox{\linewidth}{5.5cm}{\includegraphics[angle=90]{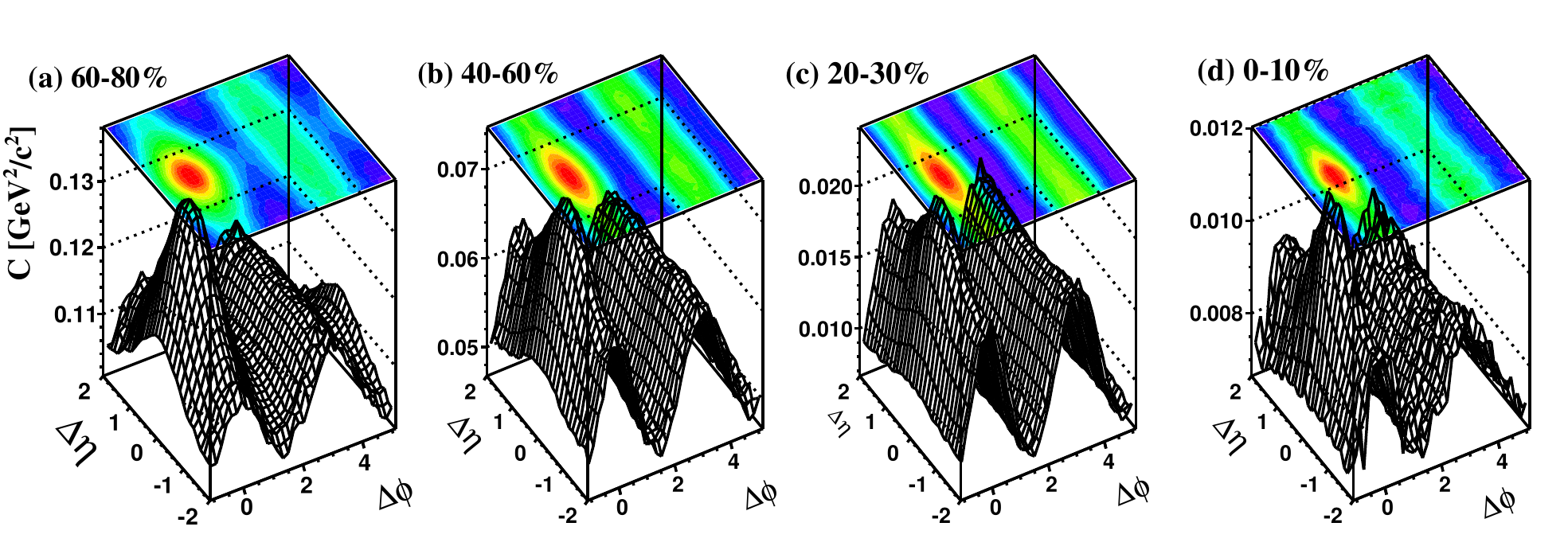}}
\caption[]{(Color online) Two-particle $p_T$ correlation function, {\it C}, plotted as a function of the particle pair pseudorapidity difference, $\Delta\eta$, and relative azimuthal angle, $\Delta\phi$ [rad], for (a) 60-80\%, (b) 40-60\%, (c) 20-30\% and (d) 0-10\% collision centralities in Au + Au interactions at $\sqrt{s_{NN}}$ = 200 \mbox{GeV} calculated with the NEXSPHERIO event generator.}
\label{CF:C}
\end{figure*}

Authors of Ref. \cite{Ref17} showed the initial density (event-by-event) fluctuations present in  NEXSPHERIO calculations produce a near-side ridge in `triggered correlation functions'. We here find that these initial density fluctuations also produce a near-side ridge in both `untriggered' and transverse momentum correlation functions. Given {\it C} has an explicit dependence on the particle momenta, one expects measurements of this variable should provide additional constraints for theoretical models.  
We further observe that the away-side ridge-like feature evolves significantly from a broad structure at $\Delta\eta \sim 0$ in most peripheral collisions to a saddle-like shape with a minimum located at $\Delta\eta \sim 0$  in the most central collisions. 

The observed build up of a near-side ridge in most central collisions is understood in NEXSPHERIO simulations to arise from hot spots extending in the longitudinal (i.e. beam) direction. Fluctuations in number and transverse position of these hot spots generate finite azimuthal anisotropy which one expects to be essentially independent of pseudorapidity in the acceptance considered in this work. Given that NEXSPHERIO involves ideal hydrodynamical transport of the energy deposited during the early moments of the collisions, one could thus anticipate that the longitudinal width of the near-side peak should be essentially unaffected by the production of the near-side ridge. 
We verify this assertion by plotting $\Delta\eta$ projections of the $R_2$ and {\it C} correlation functions in the range $|\Delta\phi|<1.0$ radians  in Figures 3 and 4 respectively. We parameterize and fit these distributions using a simple two-component ansatz to study their evolution with collision centrality.
\begin{linenomath*}
\be
 f\left( b,a ,\sigma \right) = b + a \exp \left( { - \Delta \eta ^2 /2\sigma^2 } \right) 
\label{Eq3}
\ee
\end{linenomath*} 
 
\begin{figure*}[h]
\resizebox{\linewidth}{6.0cm}{\includegraphics{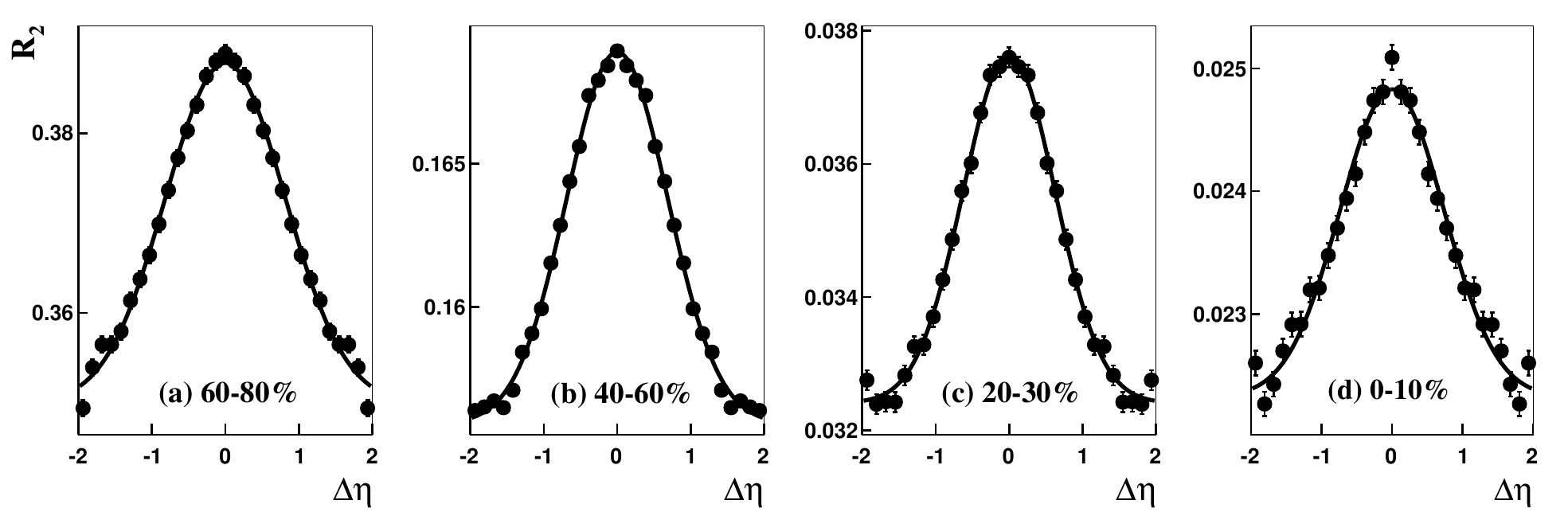}}
\label{Fig_ProjR2}
\caption[]{Projections of the correlation function $R_{2}$ for $|\Delta \phi | < 1.0$ radians on the $\Delta\eta$ axis for (a) 60-80\%, (b) 40-60\%, (c) 20-30\% and (d) 0-10\% centralities in Au + Au collisions at $\sqrt{s_{NN}}$ = 200 \mbox{GeV} from NEXSPHERIO event generator.}
\end{figure*}

\begin{figure*}[h]
\resizebox{\linewidth}{6.0cm}{\includegraphics[angle=90]{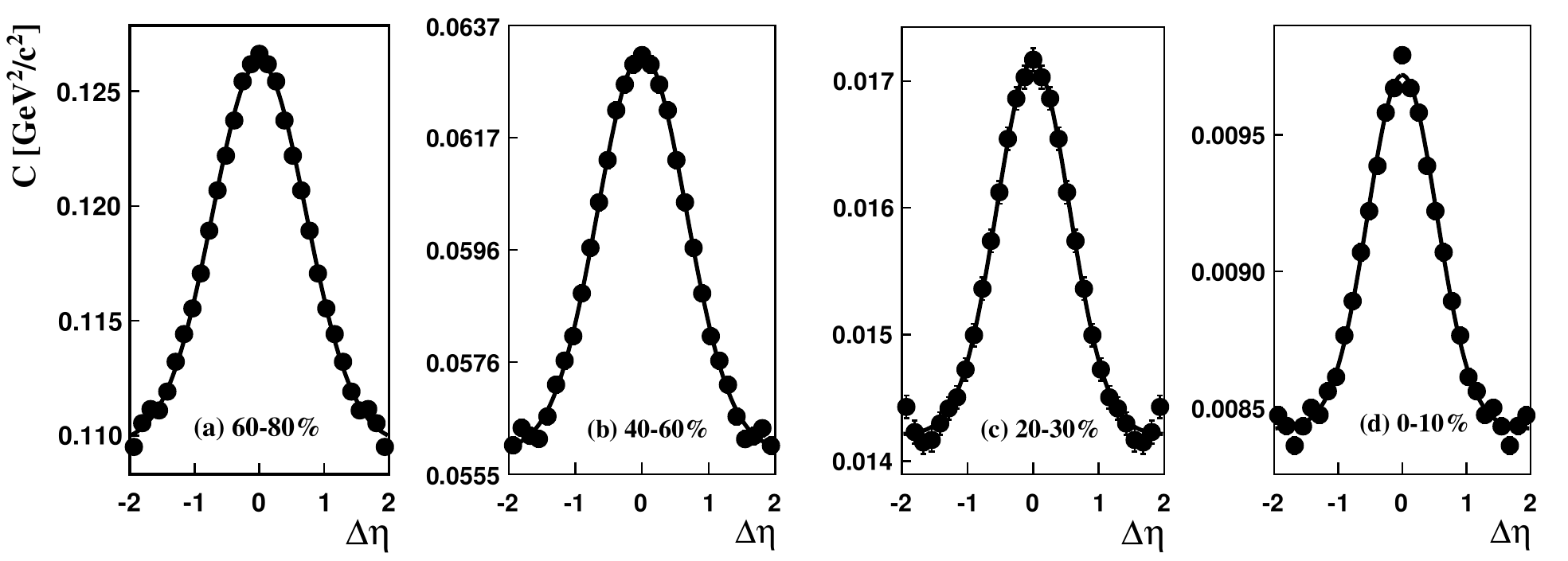}}
\label{Fig_ProjC}
\caption[]{Projections of the correlation function {\it C} for $|\Delta \phi | < 1.0$ radians on the $\Delta\eta$ axis for (a) 60-80\%, (b) 40-60\%, (c) 20-30\% and (d) 0-10\% centralities in Au + Au collisions at $\sqrt{s_{NN}}$ = 200 \mbox{GeV} from NEXSPHERIO event generator.}
\end{figure*}
 
Fit results are presented in Table I and II for observable $R_2$ and {\it C} respectively. We first observe that the near-side peak amplitude of both observables decreases monotonically with increasing collision centrality.  We also observe the emergences of
strong 
 azimuthal modulation in mid-central collisions, and the rise of a near-side ridge extending
over a long $\Delta\eta$ range in the most central collisions. These result from the formation and evolution of the medium, with initial fluctuations,  and the effects of radial and elliptic flow, embodied in the NEXSPHERIO calculations.
We furthermore find  that the distribution widths, $\sigma$, exhibit a decreasing trend with increasing collision centrality. Specifically, we find that 
the $\Delta\eta$ width of {\it C} monotonically evolves from $0.72 \pm 0.02$ units of rapidity in peripheral collisions,  to 0.54 $\pm$ 0.03 in most central collisions. $R_2$ exhibits a similar behavior, with an initial width of 0.81$\pm$ 0.02 in peripheral collisions, and reduces to 0.63 $\pm$ 0.01 in 20-30\% collisions. It however rises slightly in most central collisions. We also characterized the width of the distributions by calculating their root-mean-square (RMS) above a constant offset evaluated based on the value of the correlation function at $\Delta\eta \sim \pm 2$. The deduced RMS  values are listed in Table I and II. They are found to be in agreement, within statistical errors, with the widths,  $\sigma$, obtained from the fits. 
\begin{table}[!ht]
\caption{Various fit parameters for $R_{2}$}
\label{Table:R2}
\centering%
\begin{tabular}{cccccc}
\toprule%
Centrality      & Offset                          & Amplitude        & $\sigma$             & RMS                   & $\chi^{2}/ndf$ \\
 60-80\%       & 0.345 $\pm$ 0.001     & 0.39                 & 0.81 $\pm$ 0.02  & 0.78 $\pm$ 0.04 & 38.32/28\\
 40-60\%       & 0.156 $\pm$ 0.001     & 0.17                 & 0.69 $\pm$ 0.01  & 0.68 $\pm$ 0.03 & 41.16/28\\
 20-30\%       & 0.032 $\pm$ 0.002     & 0.038               & 0.63 $\pm$ 0.01  & 0.63 $\pm$ 0.03 & 24.27/28\\
 0-10\%         & 0.022 $\pm$ 0.002     & 0.025               & 0.74 $\pm$ 0.03  & 0.75 $\pm$ 0.04 & 37.79/28\\
 \hline
\end{tabular}
\end{table}

\begin{table}[ht]
\caption{Various fit parameters for ${\it C}$}
\label{Table:C}
\centering%
\begin{tabular}{cccccc}
\toprule%
Centrality      & Offset                              & Amplitude   & $\sigma$             & RMS                    & $\chi^{2}/ndf$  \\
 60-80\%       & 0.1096 $\pm$ 0.0002     & 0.126          & 0.72 $\pm$ 0.02  & 0.71 $\pm$ 0.04  & 23.61/28\\
 40-60\%       & 0.0559 $\pm$ 0.0001     & 0.063         & 0.67 $\pm$ 0.01   & 0.66 $\pm$ 0.03   & 16.23/28\\
 20-30\%       & 0.0142 $\pm$ 0.0001     & 0.017         & 0.56 $\pm$ 0.01   & 0.59 $\pm$ 0.03   & 17.55/28  \\
 0-10\%         & 0.0084 $\pm$ 0.0001     & 0.0097       & 0.54 $\pm$ 0.03   & 0.58 $\pm$ 0.02   & 33.99/28  \\
 \hline
\end{tabular}
\end{table} 

Figure \ref{RMSC} shows a comparison of the RMS obtained with NEXSPHERIO (black squares) with those  reported by STAR (black circles) for Au + Au collisions at $\sqrt{s_{NN}}$ = 200 \mbox{GeV}. The RMS predicted  by NEXSPHERIO substantially exceeds the value measured by STAR in most peripheral collisions. We find additionally that NEXSPHERIO 
predicts a progressive narrowing of the correlation function for increasing number of participants, at variance with the observation, by the STAR experiment, of an increasing broadening for larger number of participants ~\cite{msharma1}. The RMS reported by STAR for most central collisions is in fact almost twice as large as that predicted by the model. In addition, the smaller value of RMS observed in the STAR peripheral data could in part be due to contributions from jet particles that will yield a narrower correlation function.        
\begin{figure*}[h]
\resizebox{9.0cm}{6.0cm}{\includegraphics[angle=90]{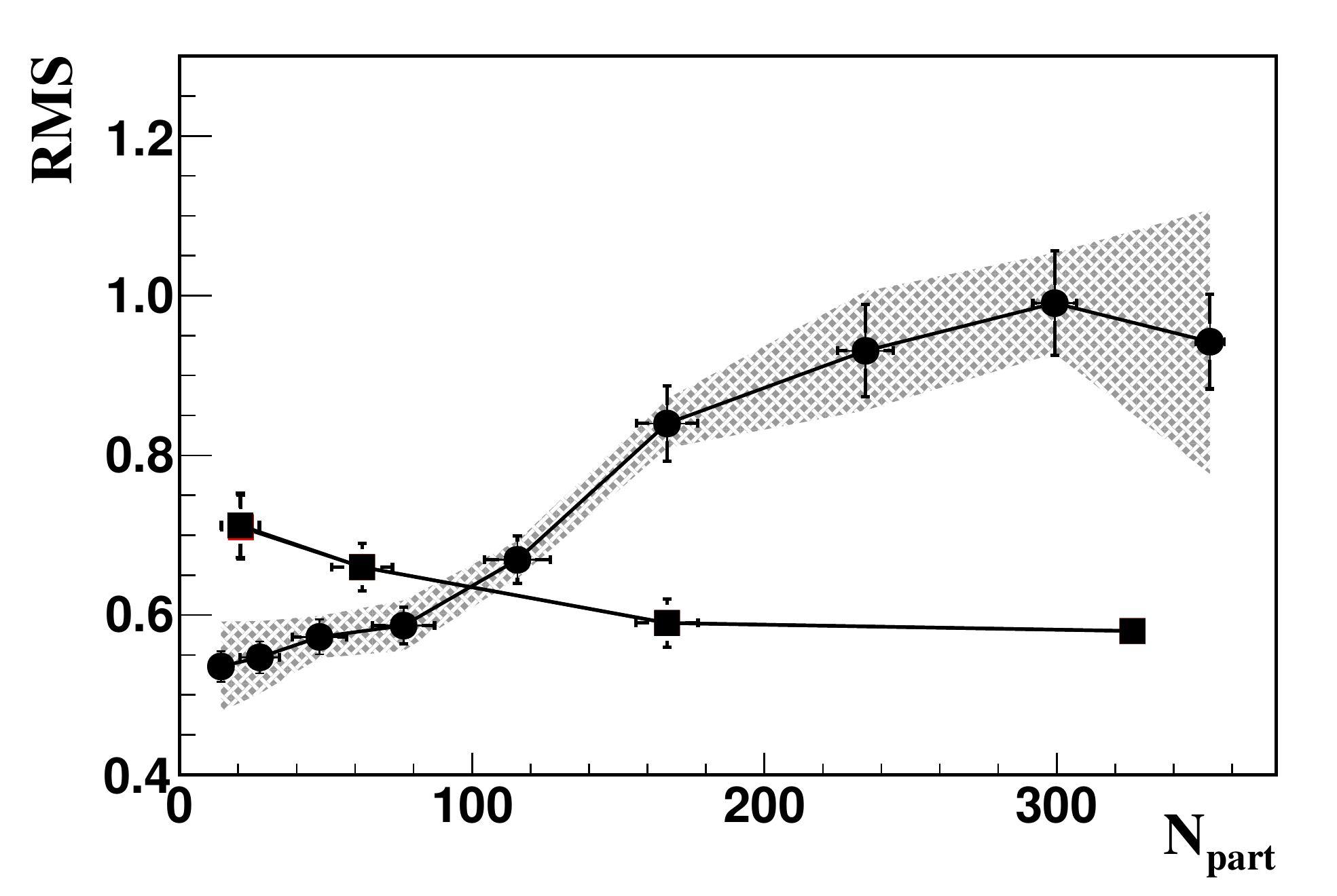}}
\caption[]{RMS of the correlation function {\it C} predicted by NEXSPHERIO (solid squares)
as a function of the number of participant nucleons,
compared  with the results reported by STAR \cite{msharma1} (solid circles), for nine centrality classes in Au+Au collisions at $\sqrt{s_{NN}}$ = 200 \mbox{GeV}.}
\label{RMSC}
\end{figure*}


We thus find that while NEXSPHERIO qualitatively reproduces the azimuthal modulation and  near-side ridge observed in experimental data it predicts a {\em narrowing} rather than the observed longitudinal broadening of the correlation function C.  
The longitudinal narrowing in NEXSPHERIO  is in large part due to the strong transverse flow. This narrowing effect of transverse flow on rapidity correlations was first discussed in \cite{voloshinRadFlow,pruneauRadFlow}. The narrowing found here is rather modest, however, amounting to approximately 0.13 units of rapidity. 

To understand why NEXSPHERIO fails to describe the measured broadening, we ask how broadening may arise in general.  In Ref.\ \cite{Gavin}, it was predicted that viscous diffusion can produce a considerable longitudinal broadening of the momentum current correlation function {\it C}.  The idea is that viscous friction acts to reduce the difference in transverse flow velocity between neighboring fluid cells. This can be measured by studying the rapidity dependence of $p_t$ fluctuations, because of the rough correspondence between spatial distance along the beam direction and rapidity. 
In principle, rapidity broadening can also result if the particle production mechanism varies appropriately with centrality. We speculate that NEXSPHERIO fails to reproduce the measured broadening because neither effects are included. 

The STAR Collaboration used their measurement of the broadening of the correlation function $C$ to estimate the viscosity per unit of entropy based on the following equation \cite{Gavin}:
\begin{linenomath*}
\begin{equation}
\sigma _c^2  - \sigma _0^2  = 4\frac{\eta}{T_{c}s} \left( {\tau _{0}^{ - 1}  - \tau _{c,f}^{ - 1} } \right)
\label{Eq:1}
\end{equation}
\end{linenomath*}
where $\sigma_c$ and $\sigma_0$ stand for the longitudinal widths of the correlation function in central collisions and at formation time, respectively. $\tau_{0}$ refers to the formation time and $\tau_{c,f}$ is  the kinetic freeze-out time at which particles have no further interactions \cite{TIME}. {\it $T_{c}$} stands for a characteristic temperature, here taken to be the critical temperature of the medium, $T_{c}=180$ MeV.
This expression neglects the fact that radial flow causes a narrowing of the correlation function $C$.  We estimate the error due to that omission using (3) as follows. A difference in the width due to flow increase the value of $\eta$ by a fraction $\Delta\eta/\eta = \Delta\sigma^2/(\sigma_c^2- \sigma_p^2)$, where $\sigma_{c,p}$ are the measured values in central and peripheral collisions. NEXSPHERIO yields a reduction of width from 0.72 in peripheral collisions to 0.54 in central collisions, while the data increases from 0.54 to 0.94, see Fig.\ 5. STAR reports a range $\eta/s = 0.06 - 0.21$ in \cite{msharma1}. We thus estimate an increase of $\Delta\eta/\eta \approx 0.38$ for the upper limit of this range. 
This suggests that  radial flow effects have in fact a somewhat limited impact on the width of the correlation function $C$ relative to that of viscous effects and can  thus be neglected, to first order, in the extraction of the fluid viscosity based on STAR data.  

\section{Conclusions}
\label{sec:conclusions}

In summary, we presented a study of the centrality dependence of the correlation functions $R_{2}$ and {\it C} in Au~+~Au collisions at $\sqrt{s_{NN}} = $~200 \mbox{GeV} based on the NEXSPHERIO model.  We find the two observables exhibit qualitatively similar shapes in $\Delta\eta$ and $\Delta\phi$, and dependence on collision centrality. Quantitative differences however arise from the explicit dependence of {\it C} on particle momenta. We find that both observables exhibit a near-side ridge in central collisions owing to event-by-event fluctuations in the initial transverse energy deposition profile. We studied near-side projections of the near-side ($|\Delta\phi|<1.0$ radians) of these correlation functions and studied their evolution with collision centrality.  We found the longitudinal width of {\it C} exhibits a small  decrease with increasing collision centrality owing to radial expansion dynamics. The magnitude of this reduction is rather modest. It is unlikely that this reduction would mask correlation broadening associated with viscous effects in real systems. Indeed, we  find that the RMS reduction observed in this work is rather modest in comparison with the sizable increase of the longitudinal width reported by STAR from their measurement of {\it C} in Au + Au collisions at $\sqrt{s_{NN}}$ = 200 \mbox{GeV}.  
From this study, it is clear that the observed increase in the width of the longitudinal correlation function {\it C} from peripheral to central collisions by the STAR experiment is not due just to variations of flow effects. In fact, our result shows that with an ideal hydrodynamic scenario, the width of the correlation function would decrease, thus suggesting that the broadening caused by other possible effects such as viscosity could be even higher than the observed values.
Taken at face value, the STAR data suggest viscous effects are indeed finite and perhaps well above the quantum bound in central Au + Au collisions.

While NEXSPHERIO does not describe the measured increase in $\Delta \eta$ with increasing centrality, the model does describe the azimuthal dependence of correlations  qualitatively well. We speculate that an extension of NEXSPERIO that includes viscous hydrodynamics, or other similar viscous hydro-models, might describe  the $\Delta \eta$ broadening reported by STAR. We surmise that this is a generic feature of models that derive long range rapidity correlations from the early time behavior followed by hydrodynamic flow  \cite{Dumitru:2010iy,LSR1,Ref12:e,LSR2,Moschelli:2009tg}. We have not dwelled on the detailed ways in which measurements of beam-energy, $p_T$, and projectile-mass  dependence might ultimately  distinguish these models. 

This work was supported in part  by U.S. NSF grant PHY-0855369 and DOE grant DE-FG02-92ER40713. We  also acknowledge support from FAPESP and FAPERJ CNPq, CAPES and PRONEX of Brazil.

\end{linenumbers}
\end{setpagewiselinenumbers}		        

\begin{thebibliography}{9}
\bibitem{Ref1} I. Arsene {\it et al.,} [BRAHMS Collaboration], Nucl. Phys. A {\bf 757} (2005) 1.
\bibitem{Ref2} B. B. Back {\it et al.,} [PHOBOS Collaboration], Nucl. Phys. A {\bf 757} (2005) 28. 
\bibitem{Ref3} J. Adams {\it et al.,} [STAR Collaboration], Nucl. Phys. A {\bf 757} (2005) 102;
\bibitem{Ref4} K. Adcox {\it et al.,} [PHENIX Collaboration], Nucl. Phys. A {\bf 757} (2005) 184.
\bibitem{Ref5} H. Zhang {\it et al.,}  Phys. Rev. Lett. {\bf 98} (2007) 212301.
\bibitem{Voloshin:2008dg} S.~A.~Voloshin, A.~M.~Poskanzer and R.~Snellings, arXiv:0809.2949 [nucl-ex].
\bibitem{Sorensen:2009cz}  P.~Sorensen, arXiv:0905.0174 [nucl-ex].
\bibitem{Shen:2011eg} C.~Shen, U.~W.~Heinz, P.~Huovinen and H.~Song, arXiv:1105.3226 [nucl-th].
\bibitem{Lacey:2010fe} R.~A.~Lacey {\it et al.},  Phys.\ Rev.\  C {\bf 82}, 034910 (2010).
 \bibitem{Kodama} T.~Kodama, T.~Koide, G.~S.~Denicol and P.~h.~Mota, Int. J. Mod. Phys. {\bf E16} (2007) 763-776.  
\bibitem{Gavin} S. Gavin and M. Abdel-Aziz, Phys. Rev. Lett. {\bf 97} (2006) 162302.
\bibitem{msharma1} M. Sharma for the STAR Collaboration, Nucl. Phys. A {\bf 830} (2009) 813c - 816c; H. Agakishiev {\it et al.,} [STAR Collaboration], arXiv:nucl-ex/1106.4334; M. Sharma and C. A. Pruneau, Phys. Rev. C {\bf 79} (2009) 024905.  
\bibitem{Ref7} J. Adams {\it et al.,} [STAR Collaboration], Phys. Rev. C {\bf 73} (2006) 064907.
\bibitem{Ref8} M. P. McCumber {\it et al.,} [PHENIX Collaboration], J. Phys. G {\bf 35} (2008) 104081.
\bibitem{Ref9} PHOBOS Correlations
\bibitem{Ref10} B. I. Abelev {\it et al.,} [STAR Collaboration], Phys. Rev. C {\bf 80} (2009) 064912.
\bibitem{Ref11} B. I. Abelev {\it et al.,} [STAR Collaboration], Phys. Rev. Lett. {\bf 103} (2009) 172301.
\bibitem{Ref12} N. Armesto {\it et al.,} Phys. Rev. Lett. {\bf 93} (2004) 172301;
\bibitem{Ref12:a} S. A. Voloshin, Phys. Lett. B {\bf 632} (2006) 490;
\bibitem{Ref12:b} M. Strickland {\it et al.,} Eur. Phys. J. A {\bf 29} (2006) 59;
\bibitem{Ref12:c} A. Majumder {\it et al.,} Phys. Rev. Lett. {\bf 99} (2007) 042301;
\bibitem{Ref12:d} E. Shuryak, Phys. Rev. C {\bf 76} (2007) 047901;
\bibitem{Ref12:e} A. Dumitru {\it et al.,} Nucl. Phys. A {\bf 91} (2008) 810;
\bibitem{Ref12:f} C. B. Chiu and R. C. Hwa, Phys. Rev. C {\bf 79} (2009) 034901.
\bibitem{Ref12:g} B. I. Abelev {\it et al.,} [STAR Collaboration], Phys. Rev. Lett. {\bf 103} (2009) 172301.
\bibitem{Ref13} S. A. Voloshin, arXiv:nucl-th/0312065.
\bibitem{Ref17} J. Takahashi, W. L. Qian, R. Andrade, F. Grassi, Y. Hama, T. Kodama, N. Xu, Phys. Rev. Lett. {\bf 103} (2009) 242301.
\bibitem{Petersen:2009vx} H.~Petersen, M.~Bleicher,  Phys.\ Rev.\  {\bf C79}, 054904 (2009). 
\bibitem{Werner:2010aa}
 K.~Werner, I.~.Karpenko, T.~Pierog, M.~Bleicher, K.~Mikhailov, Phys.\ Rev.\  {\bf C82}, 044904 (2010).
\bibitem{Holopainen:2010gz}
 H.~Holopainen, H.~Niemi, K.~J.~Eskola,
 Phys.\ Rev.\  {\bf C83}, 034901 (2011).
\bibitem{Hirano:2009bd}
 T.~Hirano, Y.~Nara,
 Nucl.\ Phys.\  {\bf A830}, 191C-194C (2009).
\bibitem{Petersen:2008dd}
 H.~Petersen, J.~Steinheimer, G.~Burau, M.~Bleicher, H.~Stocker, Phys.\ Rev.\  {\bf C78}, 044901 (2008).
\bibitem{Sorensen:2011hm}
 P.~Sorensen, B.~Bolliet, A.~Mocsy, Y.~Pandit, N.~Pruthi, [arXiv:1102.1403 [nucl-th]].
\bibitem{Alver:2010gr}
 B.~Alver and G.~Roland, Phys.\ Rev.\  C {\bf 81}, 054905 (2010) [Erratum-ibid.\  C {\bf 82}, 039903 (2010)].

\bibitem{Petersen:2011fp} H.~Petersen, C.~Greiner, V.~Bhattacharya, S.~A.~Bass, [arXiv:1105.0340 [nucl-th]].
\bibitem{Petersen:2010cw} H.~Petersen, G.~-Y.~Qin, S.~A.~Bass, B.~Muller, Phys.\ Rev.\  {\bf C82}, 041901 (2010).
\bibitem{Armesto:2006bv} N.~Armesto, L.~McLerran, C.~Pajares, Nucl.\ Phys.\  {\bf A781}, 201-208 (2007).
\bibitem{Song:2011qa}  H.~Song, S.~A.~Bass, U.~Heinz, Phys.\ Rev.\  {\bf C83}, 054912 (2011).
\bibitem{Romatschke:2007mq} P.~Romatschke, U.~Romatschke, Phys.\ Rev.\ Lett.\  {\bf 99}, 172301 (2007).
\bibitem{Ref18} Y. Hama, T. Kodama, and O. Socolowski, Braz. J. Phys. {\bf 34}, 24 (2005).
\bibitem{Ref19} W. L. Qian {\it et al.,} Braz. J. Phys. {\bf 37},  767 (2007). 
\bibitem{Ref15} H. J. Drescher {\it et al.,} Phys. Rev. C {\bf 65},  054902 (2002).
\bibitem{Ref16} C. E. Aguiar, Y. Hama, T. Kodama, and T. Osada, Nucl. Phys. {\bf A698}, 639 (2002). 
\bibitem{LUMPY} R. Andrade, F. Grassi, Y. Hama, W.-L. Qian, J. Phys. {\bf G37}, 094043 (2010). 
\bibitem{TIME} J. D. Bjorken, Phys. Rev. D {\bf 27}, 140 (1983);
			D. Teaney, Prog. Part. Nucl. Phys. {\bf 62}, 451 (2009);
			K. Dusling {\it et al.,} Nucl. Phys. A {\bf 836}, 159 (2010);
			M. Luzum and P. Romatschke, Phys. Rev. Lett. 103,  262302 (2009). 
\bibitem{Bound} G. Policastro, D. T. Son and A. O. Starinets, Phys. Rev. Lett. {\bf 87} (2001) 081601; P. K. Kovtun, D. T. Son and A. O. Starinets, Phys. Rev. Lett. {\bf 94} 111601 (2005). 
\bibitem{voloshinRadFlow}
Sergei A. Voloshin, Nucl. Phys. A {\bf 749},  287 (2005).
\bibitem{pruneauRadFlow}
Claude A.\  Pruneau, Sean Gavin, Sergei A.\ Voloshin,  Nucl.\ Phys.\ A {\bf 802}, 107 (2008).
\bibitem{Dumitru:2010iy}  A.~Dumitru, K.~Dusling, F.~Gelis, J.~Jalilian-Marian, T.~Lappi and R.~Venugopalan,  Phys.\ Lett.\  B{\bf 697}, 21 (2011).
\bibitem{LSR1} J. Liao and E. Shuryak, Phys.\ Rev.\ C {\bf 77},  064905 (2008).
\bibitem{LSR2} S. Gavin, L. McLerran and G. Moschelli, Phys.\ Rev.\ C {\bf 79}, 051902 (2009). 
\bibitem{Moschelli:2009tg} G.~Moschelli and S.~Gavin,  Nucl.\ Phys.\  A {\bf 836}, 43 (2010).
\end{thebibliography}
\end{document}